# Analyzing Currency Fluctuations: A Comparative Study of GARCH, EWMA, and IV Models for GBP/USD and EUR/GBP Pairs

**Narayan Tondapu**
narayan.tondapu@gmail.com
Microsoft, Redmond, Washington, USA

**Abstract**

In this study, we examine the fluctuation in the value of the Great Britain Pound (GBP). We focus particularly on its relationship with the United States Dollar (USD) and the Euro (EUR) currency pairs. Utilizing data from June 15, 2018, to June 15, 2023, we apply various mathematical models to assess their effectiveness in predicting the 20-day variation in the pairs' daily returns. Our analysis involves the implementation of Exponentially Weighted Moving Average (EWMA), Generalized Autoregressive Conditional Heteroskedasticity (GARCH) models, and Implied Volatility (IV) models. To evaluate their performance, we compare the accuracy of their predictions using Root Mean Square Error (RMSE) and Mean Absolute Error (MAE) metrics. We delve into the intricacies of GARCH models, examining their statistical characteristics when applied to the provided dataset. Our findings suggest the existence of asymmetric returns in the EUR/GBP pair, while such evidence is inconclusive for the GBP/USD pair. Additionally, we observe that GARCH-type models better fit the data when assuming residuals follow a standard t-distribution rather than a standard normal distribution. Furthermore, we investigate the efficacy of different forecasting techniques within GARCH-type models. Comparing rolling window forecasts to expanding window forecasts, we find no definitive superiority in either approach across the tested scenarios. Our experiments reveal that for the GBP/USD pair, the most accurate volatility forecasts stem from the utilization of GARCH models employing a rolling window methodology. Conversely, for the EUR/GBP pair, optimal forecasts are derived from GARCH models and Ordinary Least Squares (OLS) models incorporating the annualized implied volatility of the exchange rate as an independent variable.

## 1. Introduction

As defined by the [1], a volatility of a financial asset is simply how much its price changes over time. Though it doesn't precisely indicate risk, it does give a clear idea about the uncertainty in investing in the asset. Predicting exchange rate volatility is crucial for global investors. Their profits are directly affected by currency fluctuations [2]–[9]. When investing abroad, they often need to convert their domestic currency into foreign currency, exposing them to risk. High volatility can lead to either profit or loss when converting profits back. Similarly, importers and exporters are impacted by exchange rate volatility. A weaker domestic currency increases import costs, affecting profit margins and consumer prices. Conversely, a stronger domestic currency benefits importers but poses challenges for exporters. Uncertainty during high volatility periods necessitates accurate forecasts to inform business decisions. Therefore, governments should pay attention to changes in foreign money values, especially when their own money is involved. Quick and big changes in the value of money from one country to another can have serious effects on a country's money system. If a country's money becomes less valuable, it can cause prices for things people buy to go up, which affects consumers. It might also make the central bank increase interest rates to deal with the situation. Plus, when a country's money isn't worth as much, it can make paying back money owed to other countries more expensive. People who try to make money by guessing how money values will change often try to use these quick changes in financial markets. They try to make money by buying things when prices are low and selling them when prices are high, or the other way around, in short periods of time. When markets are changing a lot, there are many chances for people to make money by guessing right on how prices will change for different things like stocks, money, goods, and digital money. While there's a chance to make a lot of money with these changes, there's also a higher chance of losing money.

The significance of volatility in pricing derivatives cannot be overstated. It plays a crucial role in determining the worth of call and put options, which are valued using the Black-Scholes model. When volatility is high, option prices rise, indicating more uncertainty and potential for significant price changes. Conversely, low volatility results in cheaper options, suggesting a quieter market with less chance of major price shifts. Traders and investors need to closely watch volatility levels as they directly impact their risk and reward evaluations when trading options. Moreover, comprehending and predicting changes in volatility is essential for effectively managing risks in portfolios and optimizing trading strategies in derivative markets. Considering the extensive impact of market volatility on various sectors and market participants, it's imperative for financial institutions worldwide to find suitable models for accurately forecasting volatility in these markets. However, this task is far from simple. Asset returns are influenced by numerous factors like market conditions, economic indicators, news events, and investor sentiment, making volatility data complex and challenging to forecast. One particular trait is the clustering effect. This means that periods of high volatility tend to happen close together, followed by periods of low volatility. Another trait is asymmetry, which means that negative events in the market, like market downturns or bad news, can cause volatility to rise more quickly and by a larger amount compared to positive events. This asymmetry is often seen in stock market data, where negative price movements make the stock riskier, leading to even more volatility. Even though currencies aren't connected to a company like stocks are, studies show that this asymmetry still exists in foreign exchange data. Possible reasons for this include actions by central banks and the effects of the base currency.

Aside from these traits, volatility isn't something we can see directly. This adds another challenge when trying to make accurate forecasts because we have to pick a substitute to represent the real volatility of the asset. There's no single correct substitute; researchers often use absolute returns, squared returns, or realized volatility. Many different techniques have been developed to handle these challenges, like Historic Volatility models, Generalized Autoregressive Conditional Heteroskedasticity (GARCH) models, and options-based Implied Volatility (IV) models. While these methods are quite different, they've all been tested extensively over many years and have proven to be effective in forecasting volatility for various assets. However, there's no universal method for forecasting volatility; different models work better in different markets. Historic volatility models, like Simple Moving Averages (SMA) and Exponentially Weighted Moving Averages (EWMA), use past data on volatility to guess how volatile things might get in the future. GARCH models also do this but in a more complex way, looking at past data on volatility, noise, and past mistakes in guessing returns. They then use this to figure out how bumpy the ride might be for an asset, like a currency. Options-based IV models, on the other hand, use fancy math and observed values of financial contracts to figure out how much investors think prices will swing.

However, in case for the Great Britain Pound (GBP). It's important in the global money game because the UK is a big player in world trade and finance. But things have been rocky for the Pound lately. The UK's decision to leave the European Union, along with the mess caused by COVID-19 and the shaky government leadership, has made the Pound's future very uncertain. And because the Pound is so important, any big moves in its value can send ripples through other financial markets. Therefore, in this study, we'll check how well different guessing methods work for predicting how much the Pound's value will swing compared to other currencies. We'll focus on the daily ups and downs of the Pound against the US Dollar (USD) and the Euro (EUR) between June 15, 2018, and June 15, 2023. This period covers the UK leaving the EU and the fallout from COVID-19. We'll try out a simple EWMA model, four GARCH models (GARCH, EGARCH, GJR-GARCH, and TGARCH), and a fancy regression model using IV from the Black-Scholes Options Pricing model. While the EWMA and implied volatility models are pretty straightforward, we'll dive deeper into the GARCH models by testing out different settings and guessing methods to find the best one for predicting how bumpy the ride will be for Pound exchange rates.

The paper is as follows; the literature review is in the next section. Data and model analysis for the empirical study are provided in Section 3. The model findings are shown in Section 4, and we wrap up the paper with some conclusions and ideas for future research in Section 5.

## 2. Literature Review

The subject of predicting volatility has been extensively examined by [1] and [10]. These studies have formed the basis for much of the research conducted over the past twenty years. The authors' work extensively underscores the significance of volatility prediction, the wide range of different approaches used historically, and the ongoing discussion regarding which methods yield the best outcomes. Ultimately, their research concludes that volatility can indeed be predicted, although there is no universal solution to the challenge. However, it seems to be widely acknowledged that options-based Implied Standard Deviation (ISD), Historical Volatility, and GARCH models are among the more effective types of models.

### 2.1 General Volatility Modelling

Though simple, models like the EWMA are seen as effective and well-regarded in the literature. [11] discovered that during periods of moderate volatility, EWMA outperformed more complex models such as GARCH and Stochastic Volatility (SV) when examining assets. They tested these models on various assets, including currency pairs and stocks, consistently finding EWMA to provide the most accurate forecasts over 1 to 5 months. [12] conducted a study comparing EWMA models with GARCH-type models on Gold volatility in Ethiopia. Their GARCH models, which included additional factors like exchange rates and inflation rates, outperformed EWMA models in forecasting accuracy, with GARCH-M providing the most accurate results. The GARCH model, originating from [13], evolved to tackle volatility challenges. Models like GJR-GARCH, TGARCH, and EGARCH aim to address asymmetric returns, while IGARCH and FIGARCH focus on persistent conditional volatility among asset returns. [13] original GARCH paper didn't primarily focus on volatility forecasting but rather on modeling inflation rate uncertainty. Comparing GARCH with an ARCH model, it was found to provide a better fit and a more reasonable lag structure. Studies support GARCH models for volatility forecasting. [14] found GARCH's superiority in modeling daily price movements compared to other methods across multiple metrics. [15] demonstrated that different GARCH models perform better in varying economic environments, such as during and after financial crises. A recent study by [16] aimed to compare two methods of making predictions using GARCH, a model commonly used in financial forecasting. One method, called rolling forecasts, uses only recent data to make predictions, while the other, called expanding window forecasts, uses all available data. The study found that expanding window forecasts outperformed rolling forecasts when applied to the Asian Stock Market, as measured by metrics like Mean Absolute Error (MAE), Root Mean Square Error (RMSE), and Mean Absolute Percentage Error (MAPE). [17] conducted a study to assess the effectiveness of different implied volatility models in predicting volatility on the DAX index. They compared these models against EWMA, ARCH, and SV models. However, their findings did not conclusively determine which model yielded the most accurate forecasts overall. Nevertheless, they highlighted that the choice of model depended on the specific purpose of volatility forecasts. For monitoring Value-at-Risk, they recommended using ARCH models. Conversely, for options pricing, they found that implied volatility forecasts and stochastic volatility forecasts offered greater utility. This observation aligns with the fact that implied volatility forecasts stem directly from options pricing models. [18] examined the reliability of implied volatility forecasts concerning the S&P 100 index. They observed that implied volatility forecasts outperformed estimates based on historical data when evaluated using out-of-sample forecasts. Interestingly, despite acknowledging that implied volatility tends to overestimate volatility, they still found it to be a superior forecasting tool. [19] explored various models for predicting volatility from a Value-at-Risk perspective. They evaluated models like EWMA, GARCH (assuming normal and t-distributions), GJR-GARCH, EGARCH, and multivariate GARCH using a dataset containing UK-based assets such as government bonds, equities, and commodities, along with a combined portfolio. Traditional metrics like MSE and MAE, as well as Value-at-Risk metrics like Time Until First Failure (TUFF), were used for evaluation. They observed differences in model performance based on the asset and evaluation metric employed, noting that random walk, EGARCH, and EWMA consistently underperformed. Despite GARCH exhibiting the best fit statistically, for Value-at-Risk forecasting, simpler models like long-term mean and ARCH were found to be more effective. Recent studies have explored integrating machine learning techniques with GARCH models to improve predictions. For instance, [20] employed Artificial Neural Networks (ANNs) to enhance GARCH forecasts by incorporating additional variables. Their study focused on forecasting the volatilities of Gold, Silver, and Copper, demonstrating reduced forecast errors compared to standard GARCH. [21] utilized ANNs independently to predict S&P 500 futures contracts' daily volatility, outperforming ISD forecasts from the Barone-Adesi and Whaley model. Meanwhile, [22] applied Support Vector Regression (SVR) to enhance GARCH, achieving superior performance across various models tested on S&P 500 stocks. Their linear kernel SVR-GARCH model outperformed other GARCH variants, indicating the potential of machine learning in volatility forecasting.

**2.2 Exchange Rate Specific Volatility Modelling**

For the precise prediction of how exchange rates fluctuate, research by [23] indicates that GARCH faces challenges when compared with forecasts derived from call options' implied volatility. Their investigation focused on individual exchange rates of the Euro, Pound, Japanese Yen, and Swiss Franc against the USD from 2002 to 2012, examining both low and high volatility periods to assess forecasting performance, with implied volatility consistently outperforming GARCH. [24] found IV to be more effective than GARCH and GJR-GARCH in forecasting EUR-USD futures volatility and associated options. They utilized various models derived from the [25] framework to calculate IV, observing its substantial explanatory power over realized interest rate volatility across daily and monthly time frames. [26] assessed the effectiveness of ISD techniques in predicting the volatility of multiple currencies over different timeframes, finding notable predictive accuracy using options-implied ISD. In contrast to the prevailing preference for IV techniques, [27] demonstrated the efficacy of GARCH models in currency volatility modeling, particularly in the case of the Deutschmark against the USD. Their research indicated that GARCH consistently outperformed a large pool of ARCH-type models. Similarly, [28] highlighted the remarkable accuracy of GARCH in inter-daily volatility forecasting, focusing on the exchange rate between the German Deutschmark and Japanese Yen against the USD, showcasing the effectiveness of GARCH models in capturing foreign exchange market dynamics. [29] found that both GARCH and EGARCH models fit currency fluctuation data well, with EGARCH slightly outperforming due to its ability to avoid integrated variances.

Additionally, they noted high leptokurtic behavior in GARCH models' residuals and the importance of non-normal distributions for specific currency pairs. [30] investigated intraday fluctuations in exchange rates and interest rates to improve forecasting accuracy. They concluded that accurate forecasts require a combination of ARCH modeling, knowledge of macroeconomic announcements timing, and understanding time-of-day/day-of-week patterns, with macroeconomic announcement timing being particularly crucial.

## 3. Empirical Study

### 3.1 Data Analysis

For our study we gathered data from Yahoo Finance, which included the daily market prices of currency pairs. With the help of the Pandas package in Python, we calculated the daily returns by looking at how the closing prices changed from one day to the next. Our first step was to study the data to find any patterns. To start, we plotted the daily price of the GBP/USD pair. From Fig. 1, it's clear that the period between June 2019 and June 2023 was very turbulent for the pound. This isn't surprising, given the significant economic events during that time, like the COVID-19 pandemic and the economic uncertainty caused by government decisions. To better understand the pound's volatility, we plotted the daily returns, showing the price changes over time. Fig. 2 highlights spikes in price changes around major events. However, this graph doesn't offer detailed statistical insights except for showing that most daily price changes usually fall within a range of +/- 2%. Fig. 3 shows the absolute daily returns, but the data appears noisy, making it hard to identify consistent patterns. Additionally, [31] cautioned against using absolute returns alone to measure volatility.

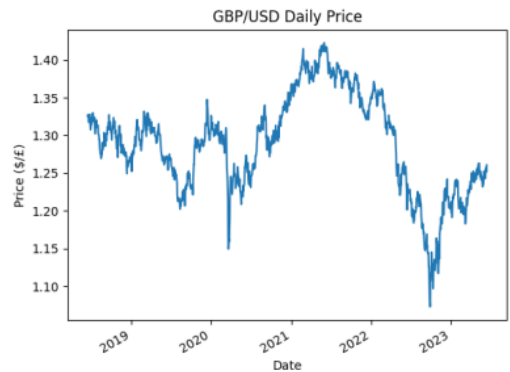

Fig. 1. Plot illustrating the daily GBP pricing versus the USD

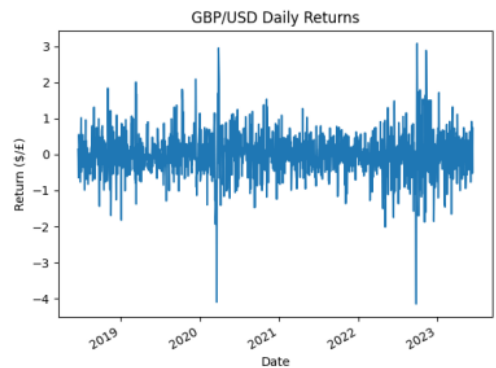

Fig. 2. Plot illustrating the daily GBP returns versus the USD

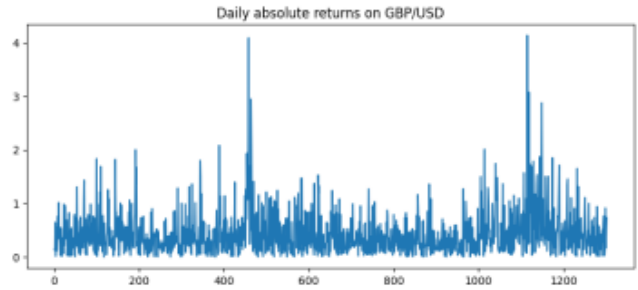

Fig. 3. The GBP/USD pair's daily returns' absolute values

To address this, Fig. 4 plots the 20-day standard deviation of returns, a common measure of volatility. This smoother trend helps us better understand volatility changes over time and will be crucial for evaluating our forecasts. Next, we explored the distribution of daily returns (Fig. 5) and analyzed its statistical properties (Fig. 6) using the Pandas library. Combining the graph and statistics, we observed a bell-shaped curve, typical of a normal distribution, with a slightly negative mean and a standard deviation. The data showed a slight negative skew, suggesting more significant negative returns than positive ones. The high kurtosis indicated a leptokurtic distribution, meaning outliers are more common, which aligns with volatility data affected by unexpected economic events. Considering these insights, we may need to explore a t-distribution as a better model for our returns data. Therefore, when testing GARCH models, we'll compare results assuming both normal and t-distributions to determine the most accurate forecasts.

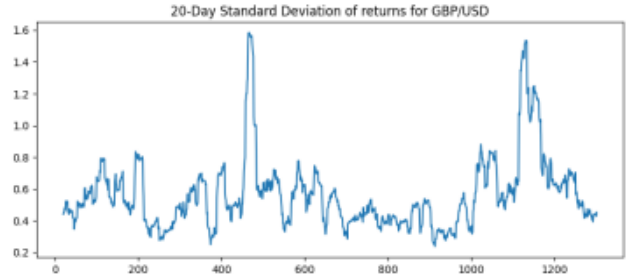

Fig. 4. The GBP/USD pair's daily returns' 20-day standard deviation

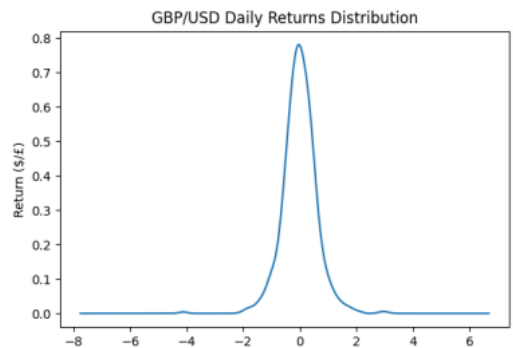

Fig. 5. The distribution of the GBP/USD pair's daily returns

```
count    1302.000000
mean       -0.002039
std         0.603512
skew       -0.188341
kurt        4.860408
min        -4.144021
25%        -0.329588
50%        -0.005832
75%         0.331654
max         3.077150
Name: Returns, dtype: float64
```

Fig. 6. Python terminal output with data on the distribution of returns in GBP/USD

We now turn our attention to examining the data concerning the GBP/EUR pair. Similarly to our analysis of the GBP/USD pair, we illustrate the daily price and the daily price change (returns) of the exchange rate. Figs. 7 and 8 depict a remarkably turbulent period from June 2019 to June 2023 for the EUR/GBP pair, mirroring the volatility observed with the GBP/USD pair.

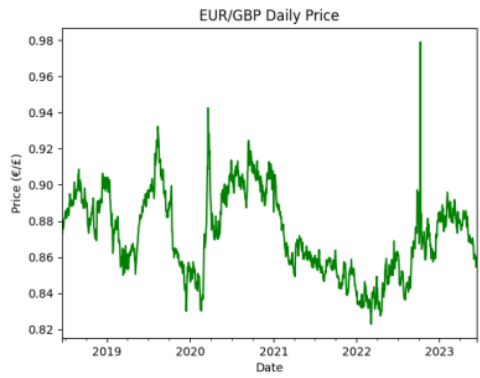

Fig. 7. Plot illustrating the daily GBP pricing versus the EUR

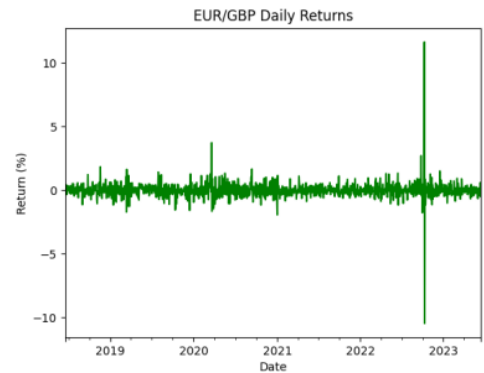

Fig. 8. Plot illustrating the daily GBP returns versus the EUR

Notably, volatility spikes occur around March 2020 and September 2022, correlating with the COVID-19 pandemic and the pound crash [32]–[38]. However, October 2022 stands out with a significant volatility spike, where the EUR value surged by over 11%. Upon graphing the data distribution (Fig. 9) and analyzing distribution statistics (Fig. 10), it becomes apparent that the dataset exhibits a notable positive skew (1.539997) and is considerably more leptokurtic (kurtosis of 147.00738). These extreme values, particularly the kurtosis, likely stem from an erroneous data point in October 2022. To rectify this issue, we smoothed out the dataset by replacing the close price with the midpoint between the close price from the previous and following days. This adjustment results in improved data presentation and more reasonable statistical properties upon re-plotting the graphs. The returns data distribution now resembles that of the GBP/USD, with a slightly higher kurtosis of 5.5467 (Fig. 11). Despite a slightly positive skew, the distribution is relatively unskewed. The data also exhibits a minor negative mean (-0.000454) and a standard deviation of 0.458229. We partitioned both datasets into in-sample and out-of-sample sets, with the last 365 observations reserved for the out-of-sample set. The in-sample set facilitates model training and evaluation of goodness of fit/statistical properties for GARCH-type and OLS models, while the out-of-sample set serves to assess the forecasting performance of all models.

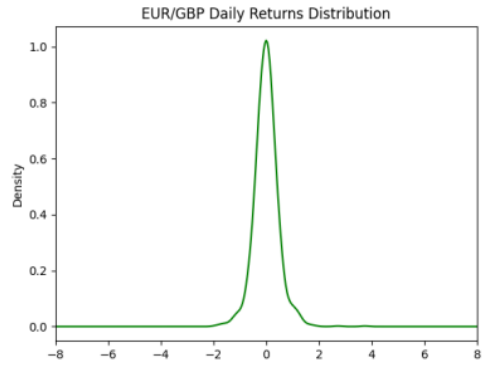

Fig. 9. The distribution of the GBP/USD pair's daily returns

```
count    1303.000000
mean        0.000481
std         0.630960
skew        1.539997
kurt      147.077338
min       -10.461696
25%        -0.252137
50%        -0.003372
75%         0.226179
max        11.640743
Name: Returns, dtype: float64
```

Fig. 10. Statistical data on the distribution of EUR/GBP returns from a Python terminal

```
count    1303.000000
mean       -0.000454
std         0.458229
skew        0.5724768398191523
kurt        5.546744336760412
min        -1.932710
25%        -0.251678
50%        -0.003515
75%         0.225068
max         3.729873
Name: Returns, dtype: float64
```

Fig. 11. Statistical data on the distribution of EUR/GBP returns from a Python terminal after cleaning the data

## 3.2 Model Analysis

### 3.2.1 Exponentially Weighted Moving Average

In our experiments, we utilized the EWMA model as the initial framework. This model, though straightforward, is highly efficient in forecasting volatility by analyzing past data trends. Its simplicity belies its effectiveness, making it a valuable tool in our research endeavors. The model's concept revolves around calculating an average of past volatility values, with more weight given to recent observations. This emphasizes the influence of recent data on future volatility. The parameter $\lambda$, which controls the weight assigned to each observation, is set at 0.97, aligning with our goal of forecasting the 20-day standard deviation of currency pairs. Utilizing such models offers the advantage of computational efficiency, enabling swift predictions of future volatility using historical data. However, the model's simplicity comes with limitations. It may not capture complex volatility patterns such as clustering and asymmetry effects, presenting a trade-off between efficiency and comprehensive data representation.

### 3.2.2 GARCH Model

The next four types of models used in the experiments all belong to the GARCH family. Specifically, GARCH, EGARCH, GJR-GARCH, and TGARCH were implemented using the arch package in Python. GARCH models are widely used for forecasting volatility. They were first introduced by [13] as an extension to the ARCH models, aiming to better capture significant fluctuations in a time series variable. Traditional ARCH models, initially proposed by [39], assume conditional heteroscedasticity, meaning that the variance of asset returns varies over time. The ARCH model treats the conditional variance of the asset as a linear function of the last $p$ squared residuals. It models the variance of a time series as a combination of a white noise component and squared residuals from previous periods. The GARCH model extends this approach by including past conditional variances into the regression equation, allowing it to capture clustering and persistence in volatility. The GARCH model, where both $p$ and $q$ parameters are set to 1, is widely considered to be the most effective for modeling asset volatility. This parameter setting includes only the white noise component, immediate lagged squared residual, and immediate lagged conditional variance in the equation (Equation 1).

$$\sigma_t^2 = \omega + \sum_{i=1}^{p} \alpha_i \epsilon_{t-i}^2 + \sum_{j=1}^{q} \beta \sigma_{t-j}^2 \tag{1}$$

Although previous literature suggests that setting $p$ and $q$ to 1 yields the best-performing GARCH model, we will systematically test different parameter combinations up to the $5^{th}$ lag using in-sample data to determine the optimal parameter sets. To assess model performance, we will utilize the Akaike Information Criterion (AIC) and Bayesian Information Criterion (BIC), which evaluate how well a model fits the dataset. The EGARCH model, stemming from [40], aims to address the inherent asymmetry in volatility trends—a challenge neglected by previous ARCH and GARCH models. These earlier models wrongly assumed volatility effects to be symmetrical, disregarding empirical evidence suggesting otherwise. EGARCH introduces three key alterations to the standard GARCH framework. First, it relaxes the non-negativity constraint on parameters $\alpha$ and $\beta$. Second, it employs the logarithm of return variance instead of variance values alone. Third, it introduces an additional coefficient, $\gamma$, to gauge the impact of the absolute residual deviation from its expected value on the logarithm of variance as shown in Equation (2).

$$log(\sigma_t^2) = \sum_{i=1}^{p} |\alpha_i \epsilon_{t-i}^2 + \gamma i(|\varepsilon_{t-i}| - E|\varepsilon_{t-i}| + \sum_{i=1}^{p} \beta_j \log(\sigma_{t-j}^2) \tag{2}$$

The GJR-GARCH model, introduced by [41], extends the GARCH model's scope by incorporating ideas from EGARCH to capture asymmetric volatility effects. This model introduces an additional parameter, allowing for differential responses to positive and negative news in financial markets. Utilizing a binary variable and an estimated coefficient, it independently models negative residuals, enhancing fit accuracy. Similarly, the Threshold GARCH model (TGARCH), as studied by [42], seeks to rectify asymmetric returns. Employing a direct regression for standard deviation by substituting squared residuals with absolute residuals, TGARCH offers an alternative approach. Like GJR-GARCH, TGARCH trains a $\gamma$ parameter to model coefficients based on the polarity of residuals, aiming to improve data fit. In estimating the GARCH model coefficients, the Maximum Likelihood Estimation (MLE) algorithm is utilized. This algorithm optimizes coefficients to maximize the log-likelihood function, assuming normal distribution of residuals. However, since returns data may better fit a t-distribution, the Quasi-MLE algorithm is often preferred, relaxing the normality assumption as shown in Equation (3).

$$\sigma_t = w + \sum_{i=1}^{q} \alpha_i [(1 - \gamma i)\varepsilon_{t-i}^+ - (1 + \gamma i)\varepsilon_{t-i}^-] + \sum_{j=1}^{p} \beta j \sigma_{t-j} \tag{3}$$

In the Python arch package, various forecasting methods are available for GARCH-type models, including rolling window and expanding window forecasts. Rolling window forecasts use the last $n$ observations to generate forecasts, while expanding window forecasts utilize all available data points. Both methods have merits and drawbacks, with expanding window forecasts expected to outperform rolling window forecasts, as indicated by [16]. To determine the most suitable forecasting methodology, both approaches will be implemented and compared.

### 3.2.3 Options Implied Standard Deviation

The approach employed by options ISD models contrasts significantly with that of the GARCH models. These models utilize options pricing frameworks like Black-Scholes and reverse-engineer them by analyzing observed option prices to compute the implied volatility of the option at a specific time. This implied volatility value reflects the market's anticipation of the future volatility of the underlying asset. The Black-Scholes model determines the price of a call option through a specific equations as shown in Equations (4), (5), and (6).

$$C_t = N(d_1)S_t - N(d_2)Ke^{-rt} \tag{4}$$

$$d_1 = \frac{\ln\frac{S_t}{K} + \left(r + \frac{\sigma^2}{2}\right)t}{\sigma\sqrt{t}} \tag{5}$$

$$d_2 = d_1 - \sigma\sqrt{t} \tag{6}$$

To predict future volatility, Options ISD is tested using an experimental procedure. Initially, annualized implied volatility data for each asset is obtained from the Bloomberg Terminal. Typically, this data is scaled down by a factor of 252, assuming 252 trading days in a year, where *n* represents the number of days for forecasting volatility. This scaled-down value serves as the forecast. However, in a study by [26], two simple OLS regression models [43] are employed. One model utilizes only the implied volatility as a parameter, while the other incorporates both implied and historical volatilities. Following this approach, we formulate regression equations to depict volatility based on the annualized data sourced from Bloomberg as shown in Equation (7).

$$\sigma_t^2 = \beta_o + \beta_1 \sigma\ implied, t-1 + \beta_2 \sigma\ actual, t-1 + \mu t \tag{7}$$

The regression model is trained using the initial 898 data points, constituting the training subset. Once trained, the model is then applied to the remaining 365 implied volatility data points, referred to as the test set. This application generates forecasts of volatility for the next 20 days, facilitating a comparative analysis with other models.

### 3.3 Evaluation Metrics

#### 3.3.1 In Sample

To assess how well various GARCH and OLS models perform on the data and select the most suitable model parameters, we will utilize the AIC and BIC metrics. These metrics help in understanding how closely the model aligns with the data while considering its complexity. The AIC is calculated as 2 times the number of model parameters minus 2 times the natural logarithm of the likelihood function, as previously described. Similarly, the BIC is derived by multiplying the number of parameters by the natural logarithm of the sample size and then subtracting 2 times the natural logarithm of the likelihood function. Both AIC and BIC scores aim to be minimized by the model. However, BIC imposes a higher penalty on the number of parameters, especially noticeable with smaller sample sizes, thus favoring simpler models over more complex ones. When determining the optimal parameter set for the GARCH models, our strategy involves selecting the model(s) with the lowest AIC and BIC scores for each model type (GARCH, GJR-GARCH, and TGARCH). We will then compare these chosen models against each other and against alternative forecasting methods for comprehensive evaluation.

#### 3.3.2 Out Sample

Understanding how distinct models fit historical data can be imperative for identifying some statistical characteristics of volatility data. Yet, the foundation principle of this study is to evaluate their performance in predicting future volatility. To assess this, specific metrics will be used such as:

- **RMSE:** The primary metric used for comparison is the RMSE (Equation 8). This metric gauges the accuracy of forecasts generated by the models in comparison to actual observed values. It's calculated using a formula where each observed data value ($y_i$) is compared with its corresponding forecast value ($\hat{y}_t$) at a given time, considering the total number of observations in the test set (*n*). RMSE is preferred over the standard MSE because it provides results on the same scale as the predicted and observed data, making the outcomes easier to comprehend.

$$R\ S\ M\ E = \sqrt{\frac{\sum(y_i - \hat{y}_t)^2}{n}} \tag{8}$$

- **MAE:** The next assessment tool we'll utilize is the MAE. This metric serves to gauge the accuracy of our forecasts. It's calculated using a simple formula (Equation 9), where '$y_i$' represents the observed data value, '$\hat{y}_t$' denotes the forecast value at a given time '*i*', and '*n*' stands for the number of observations in our test set. This method offers valuable insights into how well our predictions align with actual data, aiding in the evaluation of our forecasting models.

$$M\ A\ E = \frac{\sum y_i - \hat{y}_t}{n} \tag{9}$$

### 4. Result Analysis

#### 4.1 Exponentially Weighted Moving Average

### 4.1.1 Forecasts

The initial examination centers on the predictions generated by the EWMA models (Figs. 12 and 13). Our analysis focuses on the preceding 365 days within the dataset. From a quantitative standpoint, the model appears to produce reasonable forecasts for both currency pairs, indicating an RMSE of 0.214 and an MAE of 0.136 for the GBP/USD pair, and an RMSE of 0.146 and an MAE of 0.095 for the GBP/USD pair as shown in Table I. However, upon visualizing the forecasts, it becomes evident that the practical utility of these forecasts over a 20-day horizon is limited. Consistent with the characteristics of the EWMA model, the projected values exhibit a similar albeit slightly smoothed trend compared to the actual values. Nevertheless, this trend notably lags behind observed standard deviations, resulting in forecasts that inadequately adjust to abrupt changes in volatility. Consequently, utilizing these forecasts as a basis for financial decision-making would likely yield unfavorable outcomes. Notably, the disparity between forecasted and actual values may be less pronounced if the predictions were employed over a shorter horizon, such as one day ahead. Although the model might not anticipate sudden shifts in volatility accurately, it would adjust more swiftly once volatility stabilizes.

TABLE I
RMSE AND MAE METRICS FOR THE FORECASTS GENERATED BY THE EWMA MODEL FOR THE GBP/USD AND EUR/GBP PAIRS

|  | RMSE | MAE |
|---|---|---|
| GBP/USD | 0.214 | 0.136 |
| EUR/GBP | 0.146 | 0.095 |

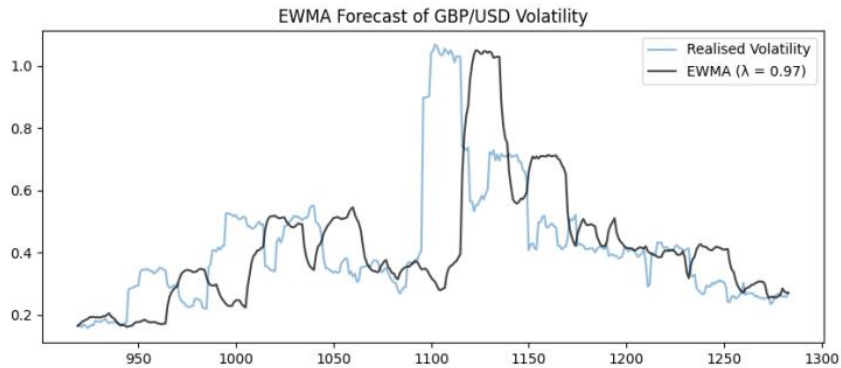

Fig. 12. For the GBP/USD pair, the EWMA forecast has $\gamma = 0.97$

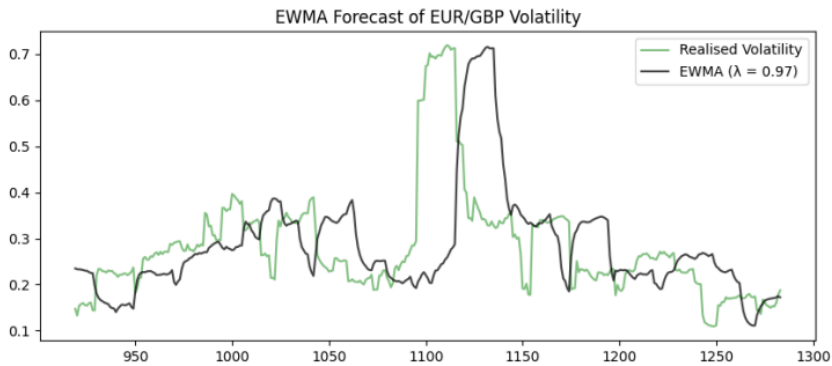

Fig. 13. For the EUR/USD pair, the EWMA forecast has $\gamma = 0.97$

### 4.2 GARCH

### 4.2.1 In Sample Fit

Our subsequent analysis focuses on examining the suitability of different parameter configurations for each GARCH model applied to the in-sample GBP/USD data. This involved generating multiple models for each GARCH specification, varying parameters ($p$ and $q$) from 1 to 5, and comparing the AIC and BIC scores. In this phase, we assume normal distribution for the residuals. The visual representations (Fig. 14 to 17) depict heatmaps showcasing AIC and BIC scores for each parameter combination tested across different model specifications. For the standard GARCH model, parameter values of $p = 2$ and $q = 2$ yield the lowest AIC score, while $p = 1$ and $q = 1$ result in the lowest BIC score. Similarly, for the EGARCH specification, optimal scores are obtained with $p = 3$ and $q = 1$ for AIC and $p = 1$ and $q = 1$ for BIC. TGARCH demonstrates its best fit with $p = 1$ and $q = 1$ for both metrics. GJR-GARCH mirrors results akin to the standard GARCH, favoring $p = 2$ and $q = 2$ for AIC and $p = 1$ and $q = 1$ for BIC. Overall, the GARCH model achieves the best AIC score, while GARCH attains the best BIC score, hinting at relatively symmetrical returns in the GBP/USD pair.

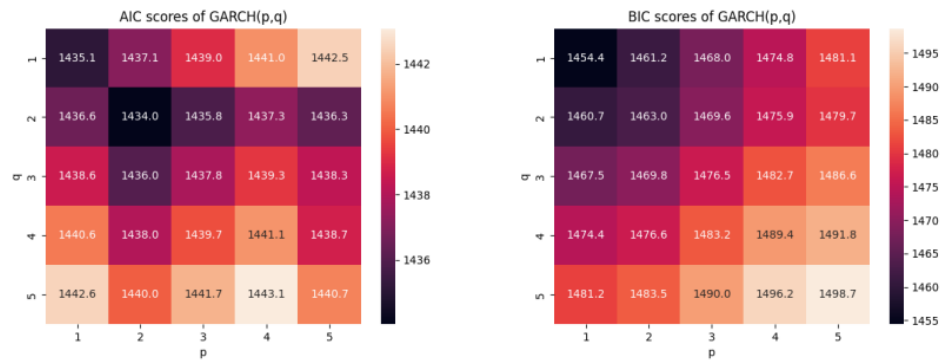

Fig. 14. Heatmap of the GARCH model parameters' AIC and BIC scores fitted to the GBP/USD pair

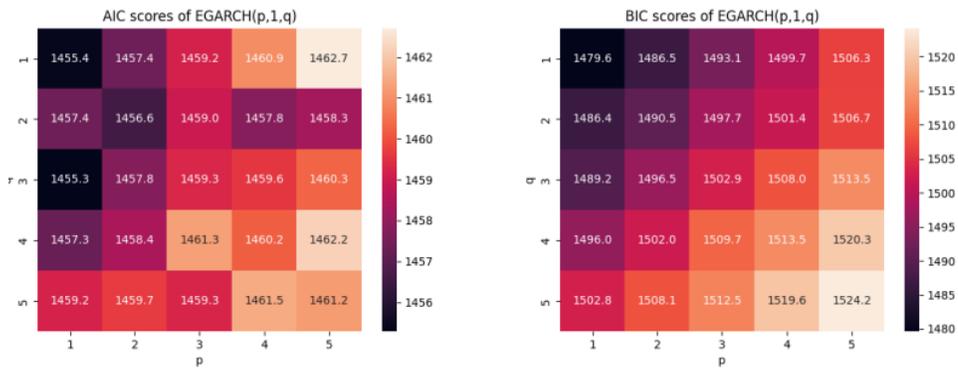

Fig. 15. Heatmap of the EGARCH model parameters' AIC and BIC scores fitted to the GBP/USD pair

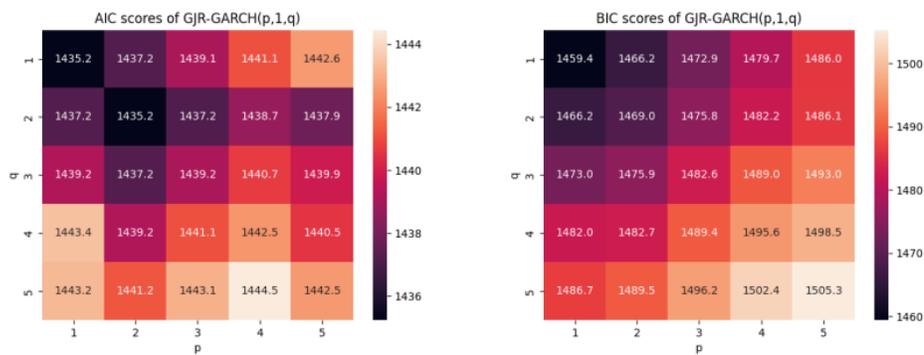

Fig. 16. Heatmap of the GJR-GARCH model parameters' AIC and BIC scores fitted to the GBP/USD pair

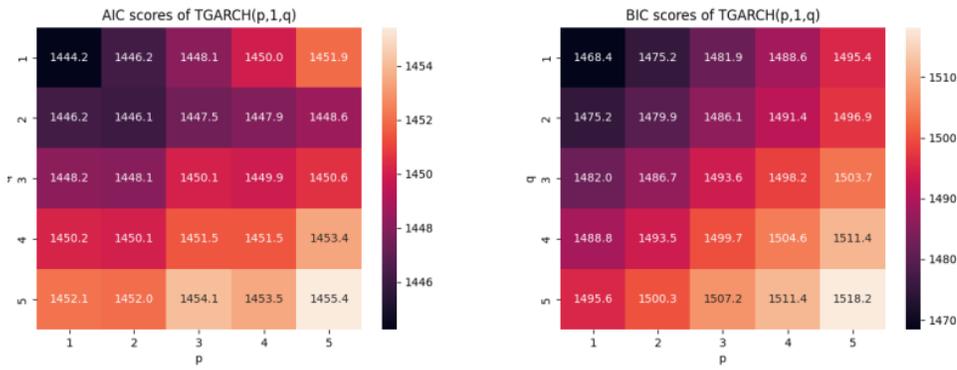

Fig. 17. Heatmap of the TGARCH model parameters' AIC and BIC scores fitted to the GBP/USD pair

The evaluation extends to the EUR/GBP pair, as detailed in Figs. 18 to 21. Here, parameter values of $p = 1$ and $q = 1$ consistently yield the optimal AIC and BIC scores across all model specifications. Specifically, TGARCH excels in AIC, while GARCH leads in BIC. This contrast from the GBP/USD results suggests potential asymmetry in the returns, particularly evident in the preference for TGARCH, known for addressing asymmetry. However, further analysis of coefficient t-statistics within the TGARCH model is required to confirm this inference. The prominence of lower lag counts in achieving superior AIC and BIC scores aligns with expectations, given the penalizing nature of both metrics towards additional variables. Notably, the prevalence of $p = 1$ and $q = 1$ across all model specifications and datasets reflects a logical choice, given the heavier penalty imposed by BIC on model complexity. There's a possibility of the data favoring a t-distribution over a normal distribution. Despite parameter determination under the assumption of a normal distribution, the arch package facilitates the transition to a t-distribution [44]–[46]. Re-estimating models under this distribution substantially reduces both AIC and BIC scores across all specifications and parameter sets, influencing future volatility forecast assumptions.

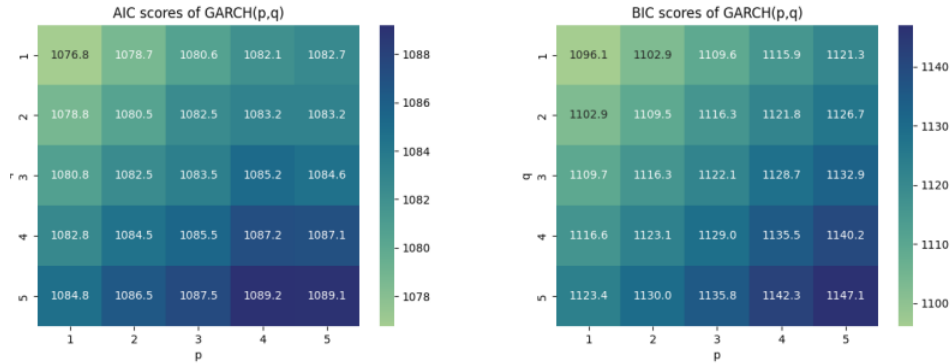

Fig. 18. Heatmap of the GARCH model parameters' AIC and BIC scores fitted to the EUR/USD pair

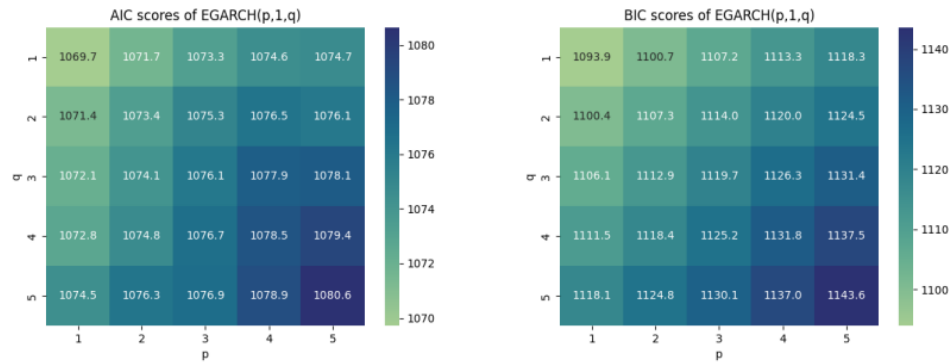

Fig. 19. Heatmap of the EGARCH model parameters' AIC and BIC scores fitted to the EUR/USD pair

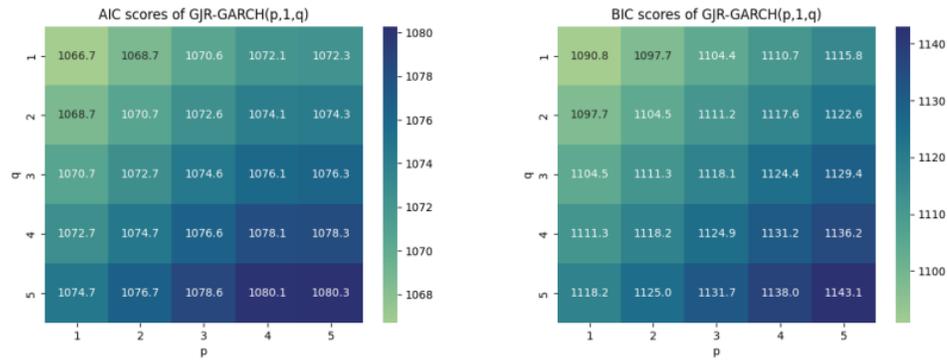

Fig. 20. Heatmap of the GJR-GARCH model parameters' AIC and BIC scores fitted to the EUR/USD pair

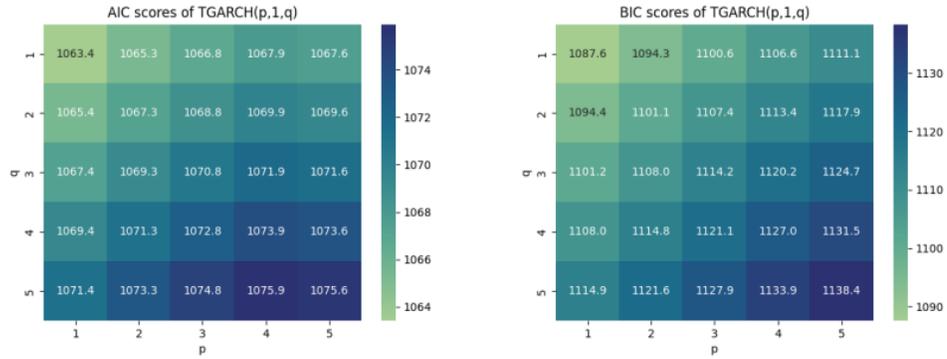

Fig. 21. Heatmap of the TGARCH model parameters' AIC and BIC scores fitted to the EUR/USD pair

The model statistics (Table II) provide further insights into each GARCH specification's fit to GBP/USD and EUR/USD returns, assuming a T-distribution. Notably, for standard GARCH models, $\alpha_1$ and $\beta_1$ coefficients exhibit strong significance, validating the model's choice. In contrast, GARCH results for GBP/USD show significant $\alpha_1$, $\alpha_2$, and $\beta_2$ coefficients, highlighting volatility persistence. However, EUR/GBP analysis suggests limitations in the appropriateness of a GARCH model. A notable observation emerges from fitting EGARCH, TGARCH, and GJR-GARCH models to GBP/USD data, where the $\gamma$ component lacks statistical significance, implying potential symmetry in returns. Conversely, for EUR/GBP, significant $\gamma$ values at various confidence intervals for EGARCH, GJR-GARCH, and GJR-GARCH (2,1,2) models indicate asymmetric returns. This contradicts prior findings by [47], highlighting potential differences in dataset characteristics and model suitability. The negative $\gamma$ values observed in GJR-GARCH and TGARCH models for EUR/GBP imply positive correlation between volatility and returns. This aligns with findings by [48], supporting the notion that volatility tends to increase when the euro appreciates against the pound, albeit requiring further validation.

TABLE II
AIC AND BIC SCORE FOR EACH MODEL WHEN ASSUMING RESIDUALS FOLLOW A T DISTRIBUTIONS

| Characteristics | Distributions | USD/GSB | | EUR/GPB | |
|---|---|---|---|---|---|
| GARCH (1,1) | t | 1444.42 | 1463.78 | 1079.99 | 1008.65 |
| | | 1409.54 | 1433.74 | 1099.36 | 1032.86 |
| GARCH (2,2) | t | 1443.23 | 1472.27 | 1083.57 | 1011.57 |
| | | 1411.08 | 1444.95 | 1112.61 | 1045.45 |
| EGARCH (1,1,1) | t | 1455.40 | 1479.61 | 1069.71 | 1005.59 |
| | | 1420.48 | 1449.54 | 1093.93 | 1034.65 |
| EGARCH (3,1,1) | t | 1455.28 | 1489.18 | 1072.15 | 1005.46 |
| | | 1418.94 | 1457.68 | 1106.05 | 1044.21 |
| GJR-GARCH (1,1,1) | t | 1444.5 | 1468.7 | 1070 1094.2 | 1005.92 |
| | | 1411.14 | 1440.18 | | 1034.96 |
| GJR-GARCH (2,1,2) | t | 1444.4 | 1478.28 | 1074 1107.89 | 1009 |
| | | 1412.39 | 1451.1 | | 1047.73 |
| TGARCH (1,1,1) | t | 1453.22 | 1477.42 | 1067.79 1092 | 1005.76 |
| | | 1419.77 | 1448.8 | | 1034.8 |

### 4.2.2 Forecasts

In this section, we utilize various statistical models, namely GARCH (1, 1), GARCH (2, 2), EGARCH (1, 1, 1), EGARCH (3, 1, 1), GJR-GARCH (1, 1, 1), GJR-GARCH (2, 1, 2), and TGARCH (1, 1, 1), to predict future volatility in the GBP/USD and EUR/GBP datasets. We base our forecasts on the assumption that the residual errors conform to a t-distribution, which provided the best fit during our model training. We commence with rolling forecasts, where we fit each GARCH-type model to the most recent 200 observations and forecast volatility for the next 20 days. Figs. 22 to 24 illustrate these rolling forecasts for the GBP/USD pair, while Figs. 25 to 27 display the same for the EUR/GBP pair. Unfortunately, we encountered optimization errors when attempting to generate rolling forecasts using the EGARCH model.

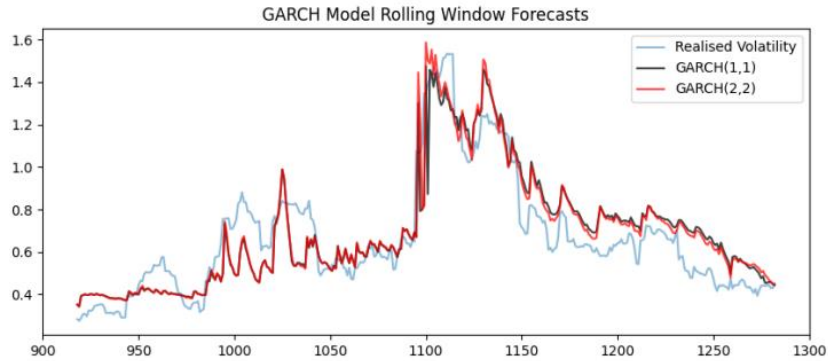

Fig. 22. Rolling estimates of the GBP/USD pair's daily returns, derived using GARCH(1,1) and GARCH(2,2)

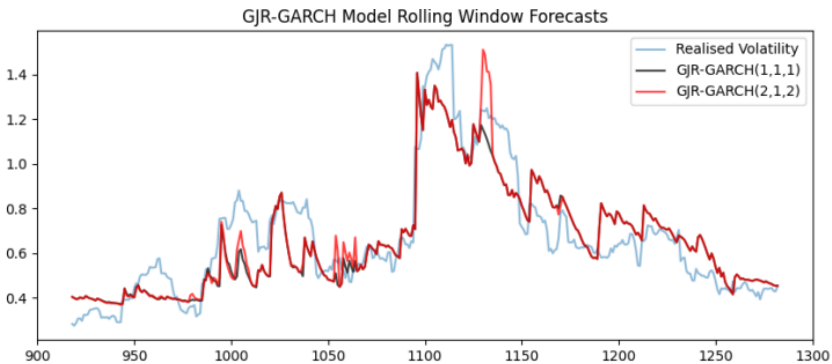

Fig. 23. Rolling predictions of the GBP/USD pair's daily returns, derived using GJR-GARCH(2,1,2) and GJRGARCH(1,1,1)

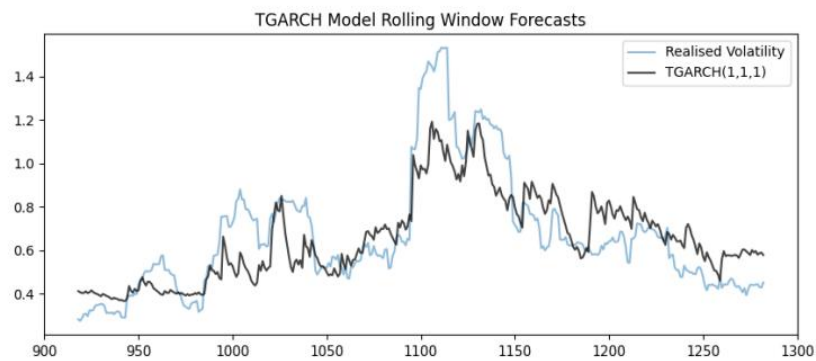

Fig. 24. Rolling estimates of the GBP/USD pair's daily returns, calculated using TGARCH(1,1,1)

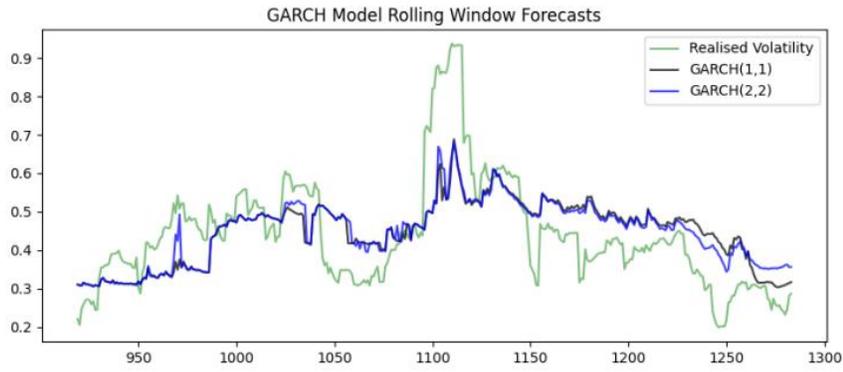

Fig. 25. Rolling estimates of the GBP/EUR pair's daily returns, derived using GARCH(1,1) and GARCH(2,2)

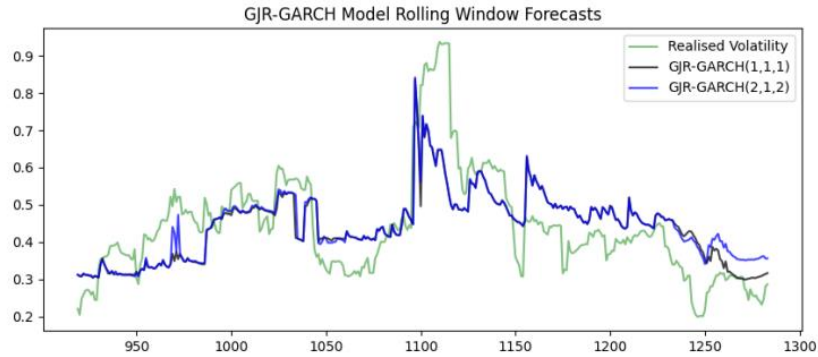

Fig. 26. Rolling predictions of the GBP/EUR pair's daily returns, derived using GJR-GARCH(2,1,2) and GJRGARCH(1,1,1)

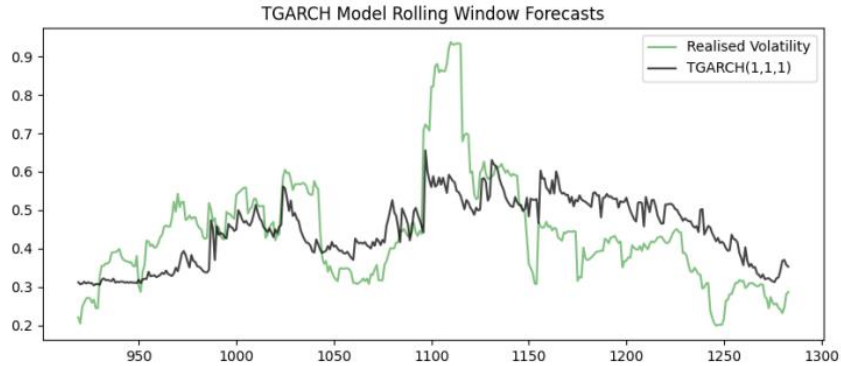

Fig. 27. Rolling estimates of the GBP/EUR pair's daily returns, calculated using TGARCH(1,1,1)

Despite GARCH models generally improving upon the forecasts generated by the EWMA model, we note some exceptions (Table III), particularly concerning the GJR-GARCH (2, 1, 2) model for GBP/USD and the TGARCH (1, 1, 1) model for EUR/GBP, which exhibit slightly higher MAE scores. Notably, the most accurate forecast for GBP/USD originates from the GJR-GARCH (2, 1 ,2) model, with an RMSE of 0.128, while for MAE, the GJR-GARCH(1, 1, 1) model performs best with a value of 0.100. Similarly, for EUR/GBP, the GJR-GARCH(1, 1, 1) model offers the most precise forecast, with an RMSE of 0.101 and an MAE of 0.080, aligning with the presence of asymmetry observed in the dataset.

TABLE III
RMSE AND MAE OF FORECASTS GENERATED BY EACH OF THE GARCH-TYPE MODELS ON THE GBP/USD AND EUR/GBP PAIRS

| Models With Rolling Expanding Forecasting Method Used | USD/GSB | | EUR/GPB | |
|---|---|---|---|---|
| | RMSE | MAE | RMSE | MAE |
| GARCH (1, 1) | 0.137 | 0.111 | 0.114 | 0.091 |
| | 0.192 | 0.134 | 0.112 | 0.087 |
| GARCH (2, 2) | 0.13 | 0.104 | 0.108 | 0.086 |
| | 0.205 | 0.142 | 0.11 | 0.085 |
| EGARCH (1, 1, 1) | N/A | N/A | N/A | N/A |
| | 0.188 | 0.126 | 0.107 | 0.084 |
| EGARCH (3, 1, 1) | N/A | N/A | N/A | N/A |
| | 0.152 | 0.105 | 0.103 | 0.080 |
| GJR-GARCH (1, 1, 1) | 0.129 | 0.1 | 0.101 | 0.080 |
| | 0.195 | 0.135 | 0.113 | 0.089 |
| GJR-GARCH (2, 1, 2) | 0.128 | 0.101 | 0.101 | 0.081 |
| | 0.205 | 0.142 | 0.111 | 0.087 |
| TGARCH (1, 1, 1) | 0.159 | 0.125 | 0.119 | 0.097 |
| | 0.189 | 0.128 | 0.11 | 0.085 |

### 4.3 Implied Volatility Model

#### 4.3.1 In Sample Fit

Fig. 28 and 29 provides insights into the statistical properties of regression models based on IV when trained on two distinct datasets. Analysis of the GBP/USD dataset reveals Model 1 attaining the highest AIC and BIC scores, contrasting with the EUR/GBP dataset where Model 2 secures the superior BIC score while Model 1 retains the leading AIC score. Interestingly, in both datasets, the coefficient $β_2$ in Model 2 exhibits a notably lower magnitude than $β_1$, accompanied by substantial standard errors, rendering it statistically insignificant at the 90% confidence level in either scenario. These findings resonate with the observations of [26], who similarly reported the insignificance of historic standard deviation coefficients. One plausible explanation for these trends is that the information encapsulated in the past 20-day standard deviation data may already be inherent within the implied volatility data. It is imperative to acknowledge that these statistics cannot be utilized to gauge the in-sample fit of these models against GARCH-type models due to their divergent predictive targets. IV regression models seek to approximate the future value of the 20-day standard deviation of returns, while GARCH-type models aim to model the conditional volatility of daily returns series.

```
Model 1
                            OLS Regression Results
==============================================================================
Dep. Variable:             target_vol   R-squared:                       0.072
Model:                            OLS   Adj. R-squared:                  0.071
Method:                 Least Squares   F-statistic:                     69.68
Date:                Tue, 29 Aug 2023   Prob (F-statistic):           2.62e-16
Time:                        16:55:50   Log-Likelihood:                 615.58
No. Observations:                 899   AIC:                            -1227.
Df Residuals:                     897   BIC:                            -1218.
Df Model:                           1
Covariance Type:            nonrobust
==============================================================================
                 coef    std err          t      P>|t|      [0.025      0.975]
------------------------------------------------------------------------------
const          0.0698      0.026      2.693      0.007       0.019       0.121
x1             0.0983      0.012      8.347      0.000       0.075       0.121
```

Fig. 28. Model 1 results on both GBP/USD and EUR/GBP

```
Model 2
                            OLS Regression Results
==============================================================================
Dep. Variable:             target_vol   R-squared:                       0.075
Model:                            OLS   Adj. R-squared:                  0.073
Method:                 Least Squares   F-statistic:                     36.14
Date:                Tue, 29 Aug 2023   Prob (F-statistic):           8.07e-16
Time:                        16:55:50   Log-Likelihood:                 616.82
No. Observations:                 899   AIC:                            -1228.
Df Residuals:                     896   BIC:                            -1213.
Df Model:                           2
Covariance Type:            nonrobust
==============================================================================
                 coef    std err          t      P>|t|      [0.025      0.975]
------------------------------------------------------------------------------
const          0.0710      0.026      2.740      0.006       0.020       0.122
x1             0.0907      0.013      7.128      0.000       0.066       0.116
x2             0.0544      0.035      1.575      0.116      -0.013       0.122
==============================================================================
Omnibus:                      407.388   Durbin-Watson:                   0.065
Prob(Omnibus):                  0.000   Jarque-Bera (JB):             2097.068
Skew:                           2.065   Prob(JB):                         0.00
Kurtosis:                       9.239   Cond. No.                         20.9
==============================================================================
```

Fig. 29. Model 2 results on both GBP/USD and EUR/GBP

#### 4.3.2 Forecasts

The regression model was utilized to predict the standard deviations of the GBP/USD and EUR/GBP pairs for the next 20 days. For GBP/USD, both IV models failed to enhance forecasting accuracy compared to the GJR-GARCH models. The RMSE and MAEs for Models 1 and 2 were 0.175, 0.111, and 0.112 respectively (Table IV), weaker than GJR-GARCH (2, 1, 2) model's RMSE of 0.128 and GJR-GARCH (1, 1, 1) model's MAE of 0.100. Interestingly, the regression model outperformed the EWMA model and expanding window forecasts of GARCH models for both RMSE and MAE. On the other hand, for EUR/GBP, regression models outperformed all GARCH variants and EWMA in terms of MAE, with Models 1 and 2 scoring 0.072 and 0.073 respectively. However, the RMSE of GJR-GARCH (1, 1, 1) remained superior to the two IV models. These findings partly contradict [26], suggesting that GARCH performs better in forecasting exchange rate volatility than implied volatility models. Despite the insignificance of $β_2$ coefficients, including them slightly improved MAE in both datasets (0.001), indicating Model 1 as the preferred regression model.

TABLE IV
RMSE AND MAE METRICS FOR THE FORECASTS GENERATED BY THE IMPLIED VOLATILITY-BASED OLS REGRESSION MODEL FOR THE GBP/USD AND EUR/GBP PAIRS

| Model | GBP/USD | | EUR/GBP | |
|---|---|---|---|---|
| | RMSE | MAE | RMSE | MAE |
| Model 1 | 0.175 | 0.112 | 0.110 | 0.073 |
| Model 2 | 0.175 | 0.111 | 0.110 | 0.072 |

## 5. Conclusions

In this investigation, we've scrutinized and evaluated various methods for predicting the volatility of the GBP/USD and EUR/GBP exchange rates. Our analysis included the EWMA model with a decay factor of 0.97, alongside the GARCH, EGARCH, GJR-GARCH, and TGARCH models, and two OLS regression models incorporating the prior day's annualized implied volatility as a predictor. By computing the 20-day standard deviations of daily returns for each currency pair, we assessed each model's capability to forecast this value 20 days ahead. To optimize the GARCH models, we iterated through $p$ and $q$ parameters from 1 to 5, selecting parameter sets yielding the best fit to the in-sample data based on AIC and BIC scores. The optimal specifications for each model type were identified as follows: GARCH (1, 1), GARCH (2, 2), EGARCH (1, 1, 1), EGARCH (3, 1, 1), GJR-GARCH (1, 1, 1), GJR-GARCH (2, 1, 2), and TGARCH (1, 1, 1). Further exploration using GARCH models revealed asymmetry in the daily returns of the EUR/GBP exchange rate, evidenced by significant values of $γ$ in EGARCH (1, 1, 1), GJR-GARCH, and TGARCH models. However, no such asymmetry was confirmed for the GBP/USD exchange rate. Additionally, modeling residual values in GARCH models with a t-distribution yielded improved fit compared to a normal distribution, as evidenced by lower AIC and BIC scores across all tested models. Evaluation of forecasting techniques between Rolling Window methods did not yield a clear preference, contrary to initial expectations. Examining the OLS regression models based on implied volatility, we found consistent significance in IV coefficients, while historical standard deviation coefficients were insignificant. Incorporating historic volatility did not significantly alter performance in out-of-sample forecasting, indicating marginal improvement in MAE. Comparing out-of-sample forecasts using RMSE and MAE metrics, GJR-GARCH models with rolling window forecasts provided the most accurate predictions for the GBP/USD pair. For the EUR/GBP pair, while GJR-GARCH models also performed well in terms of RMSE, OLS regression models outperformed in terms of MAE. Despite the strengths observed in GJR-GARCH models for GBP-related exchange rates, further research is recommended. Future investigations should extend to other currency pairs involving the pound, varying forecast horizons, and alternative evaluation metrics. Additionally, exploring recent adaptations of GARCH models integrating machine learning techniques may offer enhanced forecasting capabilities. This study acknowledges its limitations and encourages exploration of alternative models beyond GJR-GARCH and OLS regression, particularly those incorporating machine learning approaches, to potentially improve forecasting accuracy.